\documentclass[aps,prc,a4paper,nofootinbib,showpacs,showkeys,
preprintnumbers,superscriptaddress,twocolumn]{revtex4}

\usepackage{amsmath}
\usepackage{amssymb}
\usepackage{bm}
\usepackage{graphicx}
\usepackage{color}
\usepackage[english]{babel}

\begin{document}

\title{Critical Temperature of Chiral Symmetry Restoration for Quark Matter with a Chiral Chemical Potential}

\author{M. Ruggieri}\email{marco.ruggieri@ucas.ac.cn}
\affiliation{College of Physics, University of Chinese Academy of Sciences, 
Yuquanlu 19A, Beijing 100049, China}

\author{G.~X.~Peng}\email{gxpeng@ucas.ac.cn}
\affiliation{%
College of Physics, University of Chinese Academy of Sciences, 
Yuquanlu 19A, Beijing 100049, China}
\affiliation{Theoretical Physics Center for Science Facilities, Institute of High Energy Physics, Beijing 100049, China}


\begin{abstract}
In this article we study restoration of chiral symmetry at finite temperature for quark matter
with a chiral chemical potential, $\mu_5$, by means of a quark-meson model with vacuum fluctuations included.
Vacuum fluctuations give a divergent contribution to the vacuum energy, so the latter
has to be renormalized before computing physical quantities. 
The vacuum term is important for restoration of chiral symmetry
at finite temperature and $\mu_5\neq0$, therefore
we present several plausible renormalization schemes for the ultraviolet divergences
at $\mu_5\neq0$. Then we compute the critical temperature as a function
of $\mu_5$. The main result of our study is that 
the choice of a renormalization scheme affects the critical temperature; 
among the three renormalization schemes we investigate, there exists one in which the critical temperature
increases with $\mu_5$, a result which has not been found before by 
chiral model studies and which is in qualitative agreement with recent lattice data as well as with
studies based on Schwinger-Dyson equations and universality arguments.

\end{abstract}


\maketitle

\section{Introduction}
Systems with chirality imbalance, namely with a finite chiral density
$n_5=n_R-n_L$ generated by quantum anomalies, have attracted some interest in recent years.
In fact gauge field configurations with a finite winding number, $Q_W$,
can change fermions chirality according to the Adler-Bell-Jackiw anomaly~\cite{Adler:1969gk,Bell:1969ts}. 
For example in Quantum Chromodynamics (QCD)  such nontrivial gauge field configurations with
$Q_W\neq0$ are instantons and sphalerons, the latter being produced 
copiously at high temperature~\cite{Moore:2000ara,Moore:2010jd}.
In QCD the time evolution of $n_5$ is governed by the equation
\begin{equation}
\frac{dn_5}{dt} =\frac{g^2 N_f}{16\pi^2}F_a^{\mu\nu}\tilde{F}^a_{\mu\nu},
\label{eq:klq}
\end{equation}
with $F$ corresponding to the field strength tensor and $\tilde F$ its dual; 
the winding number is defined in terms of the field stregth tensor as
\begin{equation}
Q_W = \frac{g^2}{32\pi^2}\int d^4x F_a^{\mu\nu}\tilde{F}^a_{\mu\nu}.
\end{equation}
The term $F\tilde F$ is not invariant under parity transformations, ${\cal P}$,
hence currents generated by quantum anomalies are referred to as ${\cal P}-$odd currents.

Study of systems with $n_5\neq0$, and in particular of ${\cal P}-$odd currents appeared many
years ago~\cite{Vilenkin:1980fu,Vilenkin:1982pn} with applications to early universe physics;
nevertheless the possibility to observe ${\cal P}-$odd phenomena
in relativistic heavy ion collisions by the celebrated Chiral Magnetic Effect~\cite{Kharzeev:2007jp,Fukushima:2008xe}
(see~ \cite{Kharzeev:2013ffa,Kharzeev:2015znc} for reviews)
has boosted the interest for the study of a medium with a net chirality,
which spread from QCD to hydrodynamics and condensed matter systems
\cite{Son:2009tf,Banerjee:2008th,Landsteiner:2011cp,Son:2004tq,
Metlitski:2005pr,Kharzeev:2010gd,Chernodub:2015gxa,Chernodub:2015wxa,Chernodub:2013kya,Braguta:2013loa}, 
and recently has been observed in zirconium pentatelluride~\cite{Li:2014bha}.

In order to describe systems with finite chirality in thermodynamical equilibrium,
it is customary to couple the chiral chemical potential, $\mu_5$,
to its conjugated quantity namely the chiral density quark operator,
$\bar\psi\gamma_0\gamma_5\psi$, in the same manner one usually couples
the quark number density $\bar\psi\gamma_0\psi$ to the conjugated quark chemical potential $\mu$,
see~\cite{Gatto:2011wc,Fukushima:2010fe,Chernodub:2011fr,Ruggieri:2011xc,Yu:2015hym,Yu:2014xoa,
Braguta:2015owi,Braguta:2015zta,Braguta:2016aov,Hanada:2011jb,Xu:2015vna} and references therein. 
Because of quantum anomaly as well as of chirality changing processes due to finite quark condensate,
$\mu_5$ is not conjugated to a quantity which is strictly conserved; therefore usually one has to assume
the system at finite $\mu_5$ in thermodynamical equilibrium is observed on a time scale much larger than
the typical time scale of the chirality changing processes.

In the context of QCD an interesting problem is the dependence of the critical temperature for
chiral symmetry restoration, $T_c$, at $\mu_5\neq0$.
Several calculations based on chiral models predict $T_c$ to be lowered by $\mu_5$,
the slope of the line $T_c(\mu_5)$ and its curvature depending on the specific model 
used~\cite{Gatto:2011wc,Fukushima:2010fe,Chernodub:2011fr,Ruggieri:2011xc,Yu:2015hym,Yu:2014xoa}.
On the other hand recent lattice data have shown that critical temperature increases 
with $\mu_5$~ \cite{Braguta:2015owi,Braguta:2015zta}.
This behavior of $T_c(\mu_5)$ was  predicted for the first time by universality arguments in~\cite{Hanada:2011jb}
and it has also been found later by solving Schwinger-Dyson equations at finite $\mu_5$~\cite{Xu:2015vna}.
Therefore it is interesting to understand if and how
chiral models  
can reproduce at least qualitatively the dependence of $T_c$ versus $\mu_5$ found on the lattice. 

As it was discussed several years ago in~\cite{Gatto:2011wc},
and then made even more transparent in~\cite{Yu:2015hym},
the behaviour of the critical temperature as a function of $\mu_5$
in chiral models is related to a competition
among the vacuum term and the thermal fluctuations. In fact although
at zero temperature the chiral chemical potential acts as a catalyzer of chiral symmetry
breaking, close to the critical temperature
thermal fluctuations become more important at $\mu_5\neq0$, destroying the chiral condensate and
leading to a decreasing critical temperature.
The final result of the calculation was that the critical temperature decreases as $\mu_5$ increases,
even if the slope of the critical line is very small compared to that of $T_c$ versus 
the baryon chemical potential. 

In this competition mechanism among vacuum term
and thermal excitations, the role of the divergent vacuum energy is crucial. 
In previous calculations the ultraviolet cutoff $\Lambda$ appears explicitly in $T_c(\mu_5)$,
so $T_c = T_c(\mu_5,\Lambda)$. It is therefore interesting to look for a 
calculation scheme in which the dependence on $\Lambda$ is removed by renormalization
of the vacuum energy. This is the main goal of the study presented here,
in which we focus on the computation of $T_c(\mu_5)$ within a
quark-meson (QM) model with renormalization prescriptions for the vacuum energy at $\mu_5\neq0$.
We already know that renormalized vacuum energy in chiral models may affect the order
of a phase transition~\cite{Skokov:2010sf,Ruggieri:2013cya}.

We will find that among three renormalization schemes (RSs) considered there exists one,
named RS3 in the main part of the text,
which predicts the increase of $T_c$ with $\mu_5$, in agreement with the
 most recent lattice data~\cite{Braguta:2015owi,Braguta:2015zta}. What we find interesting is that
the evolution of $T_c$ with $\mu_5$ is achieved without the addition of any 
extra coupling term in the QM lagrangian, and within a simple mean field calculation.
We thus find the result a nice example of how the proper treatment of the divergent
vacuum energy in chiral models can affect the computation of physical quantities.
The qualitative agreement of RS3 with lattice data is encouraging so we are tempted to take RS3
as the best RS among the ones studied here; however we should wait for the results
of lattice simulations with masses close to the physical ones before making such a statement,
and at the moment the most fair attitude is to consider the results of the present study as an exploration of the possible
scenarios that the QM model can predict.

The plan of the article is as follows. In Section II we review briefly the QM model and the renormalization
of the vacuum energy at $\mu_5\neq0$. In Section III we extend the renormalization to the case $\mu_5\neq0$,
introducing the three RSs. In Section IV we present the main result of our study, namely the critical temperature
as a function of $\mu_5$, obtained within the three RSs. Finally in Section V we draw our conclusions.

\section{Quark-Meson model\label{Sec:QM}}
In this Section we review very briefly the quark-meson (QM) 
model~\cite{Gervais:1969zz,Mota:1999hb,Jungnickel:1995fp,Herbst:2010rf,Schaefer:2004en},
focusing in particular on the thermodynamic potential in the mean field approximation and
on the renormalization of the vacuum term at $\mu_5=0$.
The QM model is defined by the following lagrangian density
\begin{eqnarray}
{\cal L} &=&{\cal L}_{\sigma,\pi}  + {\cal L}_q,
\end{eqnarray}
where
\begin{equation}
{\cal L}_{\sigma,\pi}  =\frac{1}{2} 
\left(
\partial^\mu\sigma\partial_\mu\sigma + \partial^\mu\bm\pi\partial_\mu\bm\pi
\right)
-U(\sigma,\bm\pi),
\end{equation}
is the lagrangian density of meson fields, with $\sigma$ representing a scalar field
and $\bm\pi$ an isotriplet pseudoscalar field. The potential $U(\sigma,\bm\pi)$ which is responsible
for spontaneous breaking of the $O(4)$ symmetry as well as for classical meson excitations spectrum
is given by
\begin{equation}
U(\sigma,\bm\pi) = \frac{\lambda}{4}
\left(\sigma^2 + \bm\pi^2 -v^2\right)^2,
\label{eq:MP}
\end{equation}
which is invariant under $O(4)$ rotations in the $(\sigma,\bm\pi)$ space.  The quark content of the model
is specified by the following lagrangian density:
\begin{equation}
{\cal L}_q = \bar\psi\left[
i\partial^\mu\gamma_\mu - g(\sigma + i \gamma_5\bm\pi\cdot\bm\tau)
\right]\psi + \mu_5\bar\psi\gamma_0\gamma_5\psi,
\end{equation}
with $\psi$ being a quark field with Dirac, color and flavor indices.  In this equation
$\mu_5$ is the chiral  chemical potential, and its conjugated
quantity is named the chiral charge density, $n_5\equiv n_R - n_L$; for a system of
$N_c\times N_f$ massless fermions one has the simple relation at zero temperature:
\begin{equation}
n_5 = \frac{N_c N_f}{3\pi^2}\mu_5^3~,
\label{eq:lq1}
\end{equation}
which is formally equivalent to the relation among quark number density $n$ 
and quark number chemical potential $\mu$ for a system of massless fermions at
zero temperature, namely
\begin{equation}
n = \frac{N_c N_f}{3\pi^2}\mu^3;
\label{eq:lq2}
\end{equation}
for the case of massive fermions in Eq.~\eqref{eq:lq2} it is enough to replace
$\mu\rightarrow\sqrt{\mu^2 - m^2}$;
on the other hand between $n_5$ and $\mu_5$
is more complicated and it depends on the way one treats the divergence in the vacuum energy,
see the next Section.

Chiral symmetry breaking in the QM model
occurs because of breaking of $O(4)$ symmetry in the meson sector
down to $O(3)$ by the condensates $\langle\sigma\rangle\neq0$
and $\langle\bm\pi\rangle=0$, which is transmitted to quarks via the Yukawa coupling 
of the latter with the condensate background $g\langle\sigma\rangle\bar\psi\psi$.
We assume $v=F_\pi$ in Eq.~\eqref{eq:MP} which implies $\langle\sigma\rangle = F_\pi$, 
where $F_\pi$ corresponds to the pion decay constant in the vacuum. With this choice
the tree level mass spectrum of $\sigma$ and $\pi$ mesons is given by
\begin{equation}
M_\sigma^2 = 2\lambda F_\pi^2,~~~M_\pi^2 = 0.
\label{eq:msigmaSQ}
\end{equation}
The symmetry breaking pattern is thus equivalent to chiral symmetry breaking in QCD due to the
chiral condensate, with a massive $\sigma$ mode and a triplet of pions behaving as Goldstone modes.

The one-loop thermodynamic potential reads then
\begin{equation}
\Omega = U(\sigma,\bm\pi) + \Omega_{q} +
\Omega_T,
\label{eq:Om}
\end{equation}
where $U(\sigma,\bm\pi)$ is the chiral invariant meson potential given by Eq.~\eqref{eq:MP}, 
$\Omega_q$ corresponds to the vacuum energy,
\begin{equation}
\Omega_q = -N_c N_f\sum_{s=\pm 1} \int\frac{d\bm p}{(2\pi)^3}\omega_s,
\label{eq:Om0}
\end{equation}
with
\begin{equation}
\omega_s = \sqrt{(p+s\mu_5)^2 + m_q^2},~~~~~m_q=g\sigma;
\label{eq:omS}
\end{equation}
finally $\Omega_T$ is the thermal contribution to the quark thermodynamic potential,
\begin{equation}
\Omega_T = -2N_c N_f\sum_{s=\pm 1} \int\frac{d\bm p}{(2\pi)^3}
\log\left(1+e^{-\beta\omega_s}\right).
\label{eq:OmT}
\end{equation}
The zero temperature thermodynamic potential Eq.~\eqref{eq:Om0} is divergent in the ultraviolet; 
often this divergence is treated in an effective way by introducing a certain regularization scheme, 
let us say an ultraviolet cutoff, and treating the cutoff as an additional parameter of the model,
fixed for example by requiring that the masses of pion and sigma mesons computed within the model, 
as well as the chiral condensate, are in agreement with the measured values of these quantites.
This choice is perfectly reasonable and legittimate, since QM model (and chiral models more generally)
should be considered, at most, as low energy models for chiral symmetry breaking of QCD; 
it is thus natural that a cutoff appears in the calculations and treat this cutoff as a parameter.
However the QM model is
renormalizable, thus it is also legittimate to perform a renormalization procedure to manage the
ultraviolet divergence of the vacuum energy getting rid of the cutoff introducing and
physical renormalization point.

In order to prepare $\Omega_q$ in Eq.~\eqref{eq:Om0} for renormalization
we formally expand Eq.~\eqref{eq:Om0} in powers of $\mu_5$; we obtain
\begin{eqnarray}
\Omega_q = \Omega_0 + \Omega_5,
\end{eqnarray}
with 
\begin{eqnarray}
\Omega_0 &=& -\frac{N_c N_f}{2\pi^2}2\int p^2 dp (p^2+m_q^2)^{\frac{1}{2}},
\label{eq:Om0bare}\\
\Omega_5 &=& -m_q^2\mu_5^2\frac{N_c N_f}{2\pi^2}\int p^2 dp
(p^2+m_q^2)^{-\frac{3}{2}}\nonumber\\
&& - \mu_5^4\frac{N_c N_f}{12\pi^2},\label{eq:Om5bare}
\end{eqnarray}
and terms with higher powers of $\mu_5$ vanish when momentum integration is extended
to infinity. To regularize the divergent contributions we follow the strategy of~\cite{Frasca:2011zn}
and introduce the functions
\begin{eqnarray}
\Omega_0(s) &=& -\frac{N_c N_f}{2\pi^2}2\mu^{2s}\int p^2 dp (p^2+m_q^2)^{\frac{1}{2}-s},
\label{eq:Om0bare2}\\
\Omega_5(\xi) &=& -m_q^2\mu_5^2\mu^{2\xi}\frac{N_c N_f}{2\pi^2}\int p^2 dp
(p^2+m_q^2)^{-\frac{3}{2}-\xi}\nonumber\\
&& - \mu_5^4\frac{N_c N_f}{12\pi^2},\label{eq:Om5bare2}
\end{eqnarray}
where $s$ and $\xi$ are complex numbers; the regularized values of $\Omega_0$ and
$\Omega_5$ will be obtained by analytical continuation of $\Omega_0(s)$ and $\Omega_5(\xi)$
for $s\rightarrow 0$, $\xi\rightarrow 0$ respectively. In the above equations we have introduced
the mass scale $\mu$ which serves to leave the mass dimension of the integrals
unchanged at $s\neq0$ and $\xi\neq0$; $\mu$ will appear only in 
the arguments of logarithms and acts as an renormalization point
only in the intermediate steps of the calculation (renormalization conditions
will help to remove any $\mu-$dependence). Performing momentum integrations
and making analytic continuation to $s=0$, $\xi=0$ we find the regularized expressions
for $\Omega_0$ and $\Omega_5$, namely
\begin{eqnarray}
\Omega_0 &=& -\frac{N_c N_f}{2\pi^2}{\cal I}_1,
\label{eq:Om0bare3}\\
\Omega_5 &=& -m_q^2\mu_5^2\frac{N_c N_f}{2\pi^2}{\cal I}_2
- \mu_5^4\frac{N_c N_f}{12\pi^2},\label{eq:Om5bare3}
\end{eqnarray}
with
\begin{eqnarray}
{\cal I}_1 &=& -\frac{m_q^4}{8s}+\frac{m_q^4}{16}
\left(
-3+2\gamma_E+2\psi(-1/2)+4\log\frac{m_q}{\mu}
\right),\nonumber\\
&&
\end{eqnarray}
and 
\begin{eqnarray}
{\cal I}_2 &=& \frac{1}{2\xi}-
\left(
\frac{\gamma_E}{2}+\frac{\psi(3/2)}{2}+\log\frac{m_q}{\mu}
\right);\nonumber\\
&&
\end{eqnarray}
in the above equations $\gamma_E$ corresponds to the Euler-Mascheroni
constant and $\psi(x)$ is the di-gamma function. The ultraviolet divergences of the vacuum energy
appear as poles in Eqs.~\eqref{eq:Om0bare3} and~\eqref{eq:Om5bare3} analogously to
what happens within dimensional regularization scheme.

Equation~\eqref{eq:Om5bare3} shows that $\mu_5$ adds a further divergence in the vacuum energy,
which needs to be renormalized by a proper renormalization condition which has to be
considered as a part of the assumptions of the model.

We now briefly sketch the standard renormalization procedure of the quantum effective potential
for the QM model at zero temperature, in the case $\mu_5=0$. We emphasize here just the few
steps which are  important for the renormalization at $\mu_5\neq 0$ which will be discussed in the
following Section~\ref{Rmu5}.
We add to $\Omega_q$ two counterterms,
\begin{equation}
\delta\Omega=\frac{\delta v}{2}\sigma^2 + \frac{\delta\lambda}{4}\sigma^4
\label{eq:ct}
\end{equation}
and impose the renormalization conditions~\cite{Suganuma:1990nn,Frasca:2011zn,Ruggieri:2013cya}
\begin{equation}
\left.\frac{\partial(\Omega_q + \delta\Omega)}{\partial\sigma}\right|_{\sigma=F_\pi}=
\left.\frac{\partial^2(\Omega_q + \delta\Omega)}{\partial\sigma^2}\right|_{\sigma=F_\pi}=0,
\label{eq:RC1}
\end{equation}
which imply the vacuum expectation value of the $\sigma$ field is not affected by the one-loop corrections
and is determined by the classical potential $U$ only. Taking into account Eqs.~\eqref{eq:ct} and
~\eqref{eq:RC1} the renormalized potential is thus given by
\begin{equation}
\Omega_q^{\mathrm{ren}} =  \frac{N_c N_f}{8\pi^2}
\left(
\frac{3}{4}m_q^4 - g^2 F_\pi^2 m_q^2 + m_q^4 \log\frac{g F_\pi}{m_q}
\right).
\label{eq:O0ren}
\end{equation}
These brief remarks about renormalization at $T=0$ will be useful in the next Section
where we describe the renormalization procedure at $\mu_5\neq0$.

\section{Renormalization at $\mu_5 \neq 0$\label{Rmu5}}
Here we discuss the renormalization procedure we use for the thermodynamic potential at zero temperature
for the case $\mu_5\neq 0$. 
The renormalization condition at $T=0$ has to be considered as a part of the assumptions in the model.
The divergence in Eq.~\eqref{eq:Om5bare3} is transmitted to physical quantities; for example
assuming that $\Omega_0$ has been already renormalized, for $\mu_5\ll F_\pi$ the solution
for the $\sigma$ condensate can be found analytically 
as $\sigma=F_\pi+\delta\sigma$ and expanding in powers of $\delta\sigma$.
After a straightorward calculation and taking into account Eq.~\eqref{eq:RC1} we
obtain, to the lowest order in $\mu_5$,
\begin{equation}
\sigma=F_\pi + \mu_5^2\frac{N_c N_f}{2\pi^2}
\frac{g^2 F_\pi}{M_\sigma^2}\left(
\frac{1}{\xi} +\text{finite terms}
\right),
\label{eq:div_con}
\end{equation}
which shows that renormalization of the vacuum part with conditions  in Eq.~\eqref{eq:RC1} 
still leaves a divergent contribution to the gap equation at $\mu_5\neq0$.
Moreover by taking the second derivative of Eq.~\eqref{eq:Om5bare3} with respect to the $\sigma$ field,
which is equivalent to compute the polarization tensor of the $\sigma$ meson at zero momentum,
it is possible to show that the screening mass of this meson gets a divergent contribution
from the quark loop, namely
\begin{equation}
g^2 \Pi_s(0) = \mu_5^2\frac{g^2 N_c N_f}{2\pi^2}
\left(
\frac{1}{\xi}+\text{finite terms}
\right),
\label{eq:pis}
\end{equation}
where $\Pi_s(Q^2)$ denotes the polarization tensor of the $\sigma$ meson at zero external momentum.

The use of renormalization is not only a formal caprice: the cutoff in momentum 
integral removes fermion modes from the vacuum which instead might play some role
at finite $\mu_5$. 
We consider for example the zero temperature chiral condensate, which can be computed easily by
taking the trace of the propagator $\langle\bar\psi\psi\rangle = -i\text{Tr}[S(x,x)]$,
with the result
\begin{eqnarray}
&&\langle\bar\psi\psi\rangle = -A\int_0^\Lambda p^2 dp
\left(
\frac{1}{\omega_+} +
\frac{1}{\omega_-}
\right),
\end{eqnarray}
where $\omega_\pm$ are defined in Eq.~\eqref{eq:omS} and $A>0$ is a numerical constant which is not important for the discussion. In the above equation we have left an explicit ultraviolet cutoff $\Lambda$ in the momentum integral.
For $\mu_5=0$ the momentum region in which the term in parenthesis takes its largest contribution
is $p\ll m_q$; on the other hand $\mu_5$ shifts gradually this domain to higher values of $p\approx\mu_5$,
as also noticed in~\cite{Braguta:2016aov},
meaning that to manage properly the chiral condensate one has to include higher momenta
and this is feasible either introducing an explicit dependence of $\Lambda$ on $\mu_5$,
or performing the renormalization of the ultraviolet divergence.

We will consider here three different renormalization schemes: the first one,
which we name renormalization scheme 1 (RS1 in the following)
where we assume the fermion vacuum term does not shift the vacuum expectation value of the $\sigma$
field at $\mu_5\neq 0$, in analogy to what we assume in the $\mu_5=0$ case; 
also the squared mass of the $\sigma$ meson is assumed to be the one at $\mu_5=0$.
The RS1 is certainly the cheapest scheme.
On the other hand we will consider also less cheap renormalization schemes,
named RS2 and RS3,
in which we add a further counterterm at $\mu_5\neq0$ and impose one renormalization
condition more with respect to RS1. We will show similarities as well as differences 
in the phsyical content of the several renormalization schemes, then focusing on the effect of the 
choice of a specific one on the restoration of chiral symmetry at finite temperature.

\subsection{Renormalization Scheme 1}
We first discuss RS1. As in the case $\mu_5=0$ 
we add to $\Omega_q$ two counterterms as in Eq.~\eqref{eq:ct}
and impose the renormalization conditions in Eq.~\eqref{eq:RC1}
which imply, among other things, that $\langle\sigma\rangle$ does not take
contributions from $\mu_5\neq0$ at $T=0$. 
The renormalized $\Omega_5$ thus reads
\begin{eqnarray}
\Omega_5^{\mathrm{ren}} &=& \mu_5^2\frac{g^2N_c N_f}{2\pi^2}\sigma^2\log\frac{\sigma}{F_\pi}
-\mu_5^2\frac{g^2 N_c N_f}{8\pi^2F_\pi^2}\sigma^4
\nonumber\\&& -\frac{N_c N_f}{12\pi^2}\mu_5^4.
\label{eq:ren_O0_mu5}
\end{eqnarray}

The renormalized potential in the Eq.~\eqref{eq:ren_O0_mu5} shows an explicit and finite dependence on 
$\mu_5$. It is worth noticing that the chiral density $n_5 = -\partial\Omega/\partial\mu_5$ 
is also finite and non-vanishing: at the global minimum of the thermodynamic potential
$\sigma=F_\pi$, which is always satisfied
at $T=0$ within RS1, we obtain indeed
\begin{equation}
n_5 = \frac{N_c N_f}{\pi^2}\left(
\frac{\mu_5^3}{3} 
+  \frac{g^2 F_\pi^2}{4}\mu_5\right).
\label{eq:N5ren}
\end{equation}
For completeness we also write down the expression for the susceptibility of the
chiral density $\chi_5\equiv - d^2\Omega/d\mu^5$:
\begin{equation}
\chi_5 = \frac{N_c N_f}{\pi^2}\left(
\mu_5^2 
+  \frac{g^2 F_\pi^2}{4} \right).
\label{eq:N5ren_}
\end{equation}
Equation~\eqref{eq:N5ren} shows the increase of $n_5$ with increasing $\mu_5$ as expected:
adding $\mu_5$ injects chirality in the system;
nevertheless within RS1 there is no effect of such chiral density on
the breaking of the $O(4)$ spontaneous symmetry breaking because of conditions~\eqref{eq:RC1}. 
Therefore within RS1 the only effect of injecting chirality at $T=0$
is a finite and condensate dependent shift of the vacuum energy.

\subsection{Renormalization Scheme 2}
The previous renormalization scheme might be satisfactory from a formal point of view; 
it seems however somehow hard to accept physically because we would expect the
quark loop at finite $\mu_5$ in the grand potential to give a contribution to the $\sigma$ condensate
and/or to the screening mass of the $\sigma$ meson,
in the same way it gives a contribution at finite temperature.
We therefore are tempted to check what is the 
effect of tilting the renormalization conditions in order to have a dependence of the
$\sigma-$condensate and/or $\sigma$ meson screening mass 
at finite $\mu_5$ and $T=0$. 

To this end we introduce renormalization scheme 2 (RS2) which
differs from RS1 because we introduce independent counterterms for 
the $\mu_5=0$ and $\mu_5\neq0$ parts of the vacuum energy.
The idea is as follows: the chiral chemical potential introduces in the vacuum energy
only one log-type divergence proportional to $\mu_5^2\sigma^2$, and this suggests the need of
one single counterterm to renormalize the divergence. Therefore to the total 
vacuum energy we add the counterterms
\begin{eqnarray}
\delta\Omega&=&\frac{\delta v}{2}\sigma^2 + \frac{\delta\lambda}{4}\sigma^4
+\frac{\gamma}{2}\mu_5^2 \sigma^2
\label{eq:ct_RS3}\\
&\equiv & \Omega_{c.t.} +\frac{\gamma}{2}\mu_5^2 \sigma^2,
\label{eq:ct_RS4}
\end{eqnarray}
instead of Eq.~\eqref{eq:ct}. The explicit $\mu_5^2$ in the above equation is just a matter of convention.
We use $\Omega_{c.t.}$ in Eq.~\eqref{eq:ct_RS4} to renormalize the $\mu_5=0$
potential, assuming the renormalization conditions
\begin{eqnarray}
&&\left.\frac{\partial(U + \Omega_{c.t.} + \Omega_0)}{\partial\sigma}\right|_{\sigma=F_\pi}=0,
\label{eq:123}\\
&&\left.\frac{\partial^2(U + \Omega_{c.t.} + \Omega_0)}{\partial\sigma^2}\right|_{\sigma=F_\pi}=
M_\sigma^2,
\label{eq:124}
\end{eqnarray}
with $M_\sigma^2=2\lambda F_\pi^2$.
Conditions~\eqref{eq:123}
and~\eqref{eq:124} are formally identical to the ones used in the previous Section to renormalize the
vacuum energy at $\mu_5=0$: they are thus enough to remove quadratic and log the divergences
appearing in $ \Omega_0$.

We now use the $\gamma-$counterterm in Eq.~\eqref{eq:ct_RS4}
to renormalize the divergence in $\Omega_5$ in Eq.~\eqref{eq:Om5bare3} . To this end
we assume the renormalization condition
\begin{equation}
\left.\frac{\partial(\Omega_5 + \gamma\mu_5^2\sigma^2/2)}{\partial\sigma}\right|_{\sigma=F_\pi}=0.
\label{eq:finalmente1}
\end{equation}
The above condition is enough to remove the divergence of $\Omega_5$:
\begin{equation}
\Omega_5^{{\mathrm{ren}}} = \mu_5^2\frac{g^2 N_c N_f}{2\pi^2}\sigma^2
\left(
-\frac{1}{2}+\log\frac{\sigma}{F_\pi}
\right) - \frac{N_c N_f}{12\pi^2} \mu_5^4,
\label{eq:f7}
\end{equation}
and together with conditions~\eqref{eq:123} and~\eqref{eq:124} it implies that
$\sigma=F_\pi$ at $T=0$: the chiral chemical potential within RS2 does not
shift the expectation value of $\sigma$. However the second derivative of $\Omega_5^{ren} $ 
with respect to $\sigma$ is finite, therefore the screening mass of the $\sigma$ meson gets
a finite $\mu_5$ dependence given by
\begin{equation}
m_\sigma^2 = M_\sigma^2 + \frac{g^2 N_c N_f}{\pi^2}\mu_5^2. 
\end{equation}
For completeness we compute also the chiral density and its susceptibility within RS2
at zero temperature; we find
\begin{equation}
n_5 = \frac{N_c N_f}{\pi^2}\left(
\frac{\mu_5^3}{3} 
+  \frac{g^2 F_\pi^2}{2}\mu_5\right),
\label{eq:N5ren2}
\end{equation}
and 
\begin{equation}
\chi_5 = \frac{N_c N_f}{\pi^2}\left(
\mu_5^2 
+  \frac{g^2 F_\pi^2}{2} \right).
\label{eq:N5ren2_}
\end{equation}

Before going ahead we would like to comment that we have checked that
instead of $\sigma=F_\pi$ in Eqs.~\eqref{eq:123},~\eqref{eq:124} and~\eqref{eq:finalmente1}
we can use $\sigma=S$ with any real value of $S$ and still the renormalization procedure works.
This opens the possibility to add by hand in RS2 a dependence $\sigma(\mu_5)$ which is certainly feasible
but it should rely on the choice of an ansatz: we prefer to rely not on an ansatz hence we 
consider in this study only the solution $\sigma=F_\pi$ within RS2.

It is also worth noticing that the chiral condensate $\langle\bar\psi\psi\rangle= \partial\Omega/\partial m_0$
has a $\mu_5$ dependence within RS2 as well as RS1 at $T=0$ even if
the $\sigma$ condensate does not depend on $\mu_5$. 
Since we are interested here only to a qualitative behaviour of  $\langle\bar\psi\psi\rangle$
versus $\mu_5$ we limit ourself to a phenomenological treatment of the divergence and introduce 
an ultraviolet cut off $\Lambda$  in the relevant momentum integrals. 
Of course this naive procedure does not affect at all the results of this Section, 
and we use it only with the purpose to illustrate qualitatively the behaviour of the chiral condensate
at finite $\mu_5$.
A straightforward calculation shows that
\begin{eqnarray}
\langle\bar\psi\psi\rangle &=& 
\langle\bar\psi\psi\rangle_0
+\mu_5^2m_q\frac{N_c N_f}{2\pi^2}
\left(
3 - 2\log\frac{2\Lambda}{m_q}
\right).
\label{eq:cc2}
\end{eqnarray}
where $\langle\bar\psi\psi\rangle_0$ is the chiral condensate
at $\mu_5=0$, $m_q$ is the solution of the gap equation and
$\Lambda$ is an ultraviolet cutoff, $\Lambda\gg m_q$.
We notice that 
the $\mu_5$-dependent term in the above equation shows
an explicit log-type divergence; the same kind of divergence appears
in full QCD using the lattice regularization~\cite{Braguta:2015owi,Braguta:2015zta}.
Equation~\eqref{eq:cc2} shows that the chiral chemical potential increases the magnitude
of the chiral condensate, its correction being negative for $\Lambda\gg m_q$. 
This is a clear evidence that $\mu_5$ favours chiral symmetry breaking at $T=0$. 
The catalysis at small temperature with $\mu_5\neq0$ has been found for the first time in~\cite{Fukushima:2010fe}
using an effective model with a coupling to the Polyakov loop~\cite{Fukushima:2003fw};
then it has also been found in several other model calculations~\cite{Gatto:2011wc,Yu:2015hym,Ruggieri:2011xc};
recently this catalysis at zero temperature has been described very nicely
in terms of BCS-like instability of the Fermi surface at $\mu_5\neq0$~\cite{Braguta:2016aov}.
Chiral condensate has been measured recently on the lattice at finite $\mu_5$
and finite temperature, and it has been found that also at temperatures below the critical temperature
$\langle\bar\psi\psi\rangle$ increases with $\mu_5$ for the case of staggered fermions~\cite{Braguta:2015zta}, 
thus supporting the catalysis of chiral simmetry
breaking at finite $\mu_5$ in qualitative agreement with Eq.~\eqref{eq:cc2}.

\subsection{Renormalization Scheme 3}
Next we turn to renormalization scheme 3 (RS3) which sligthly differs
from RS2 because it allows a dependence of the $\sigma$ condensate on $\mu_5$
at $T=0$. In fact within RS2 the effect of $\mu_5$ at zero temperature is a shift
of the screening mass of the $\sigma$ meson, at the same time assuming $\sigma=F_\pi$
for any value of $\mu_5$. Once again this choice is certainly satisfactory from a formal 
point of view, but probably still lacks some physical content because if $\mu_5$ acts
as a catalyzer of spontaneous chiral symmetry breaking, then we would expect the $\sigma$
condensate to increase with $\mu_5$. 

Within RS3 we take counterterms as in RS2, see Eqs.~\eqref{eq:ct_RS3} and~\eqref{eq:ct_RS4}
and assume the conditions~\eqref{eq:123} and~\eqref{eq:124}.
On the other hand, instead of condition~\eqref{eq:finalmente1} which eventually constraints
$\sigma=F_\pi$ for any $\mu_5$, we assume the less severe condition
\begin{equation}
\left.\frac{\partial^2(\Omega_5 + \gamma\mu_5^2\sigma^2/2)}{\partial\sigma^2}\right|_{\sigma=F_\pi}=0.
\label{eq:finalmente2}
\end{equation}
The above equation is equivalent to a minimal subtraction scheme for the
polarization tensor of the $\sigma$ meson, in which the dressed propagator
for this quantum fluctuation at $T=0$ is given by
\begin{equation}
D^{-1}_\sigma(Q^2) = Q^2 - M_\sigma^2 
+g^2 \left[\Pi_s(\sigma,Q^2) - \Pi_s(F_\pi,0)\right],
\label{eq:boh}
\end{equation}
where $\Pi_s$ corresponds to the polarization tensor; within RS3 whenever in the
ground state $\sigma=F_\pi$ then the screening mass of the $\sigma$ meson
does not take contribution from the zero temperature part of the quark loop.

The above conditions are enough to remove the divergence of $\Omega_5$
which in RS3 reads
\begin{equation}
\Omega_5^{\mathrm{ren}} = \mu_5^2\frac{g^2 N_c N_f}{2\pi^2}\sigma^2
\left(
-\frac{3}{2}+\log\frac{\sigma}{F_\pi}
\right) - \frac{N_c N_f}{12\pi^2} \mu_5^4.
\label{eq:f8}
\end{equation}
The present renormalization scheme is quite interesting because it permits the computation
of the $\mu_5$ dependence of the $\sigma$ condensate at $T=0$, as well as
the shift of the screening mass of the $\sigma$ meson. Although the previous
equations are valid for any value of $\mu_5/\sigma$ we limit ourselves to a solution
of the gap equation for the $\sigma$ condensate for small values of $\mu_5/F_\pi$
because in this case we obtain simple analytical relations, and more important the role
of the vacuum term on the critical temperature will be more transparent.

The $\sigma$ condensate in the limit $\mu_5/F_\pi\ll1$
is computed by minimizing $\Omega=U+\Omega_0^{ren} + \Omega_5^{ren}$
with respect to $\sigma$, with $U$ and $\Omega_0^{ren}$ defined in 
Eqs.~\eqref{eq:MP} and~\eqref{eq:O0ren}, assuming the ansatz $\sigma=F_\pi(1+\delta)$
with $\delta\ll 1$ and expanding to the lowest non trivial order in $\delta$. The calculation
is straightforward and not informative  so we limit ourself to quote the final result, namely
\begin{equation}
\sigma = F_\pi 
\left(
1 + \frac{\mu_5^2}{F_\pi^2}\frac{g^2 N_c N_f}{\pi^2}\frac{F_\pi^2}{M_\sigma^2}
\right),~~~\text{RS3}.
\label{eq:dpo}
\end{equation}
The above equation shows the $\sigma$ condensate increases with increasing $\mu_5$
as expected. We can also compute the shift of the screaning mass of the $\sigma$ meson
by computing the second derivative of $\Omega$ with respect to $\sigma$ at the
global minimum of the potential~\eqref{eq:dpo}. We find
\begin{equation}
m_\sigma^2 = M_\sigma^2 +
\frac{\mu_5^2}{M_\sigma^2}\frac{g^2 N_c N_f}{\pi^2}
\left( 
3 M_\sigma^2 
-\frac{g^4 N_c N_f F_\pi^2}{\pi^2}
\right).
\end{equation}
For completeness we report the zero temperature relation among chiral density and chemical potential
within RS3 which is easily obtained from Eq.~\eqref{eq:f8}, namely
\begin{eqnarray}
n_5 &=&  \frac{N_c N_f}{\pi^2}\left(
\frac{\mu_5^3}{3} 
+  \frac{3g^2 F_\pi^2}{2}\mu_5\right) ;
\end{eqnarray}
we also compute the zero temperature chiral density susceptibility, whose calculation is elementary but requires
some care because the condensate within RS3 depends on $\mu_5$ and this dependence
has to be taken into account when the second derivative of the vacuum energy with respect
to $\mu_5$ is performed: 
\begin{eqnarray}
&&\chi_5 =\frac{N_c N_f}{\pi^2}\left[ 
\mu_5^2
\left(1
+\frac{6g^4 F_\pi^2}{M_\sigma^2} \frac{N_c N_f}{\pi^2}
\right)+\frac{3g^2 F_\pi^2}{2}\right] .\nonumber\\
&&
\end{eqnarray}

For sake of reference in Table~\ref{tab:1} we summarize the physical content of the three renormalization schemes
discussed above. We have not specified the values of the positive proportionality constants
because the only purpose of the Table is to collect the relevant similarities and differences of the 
three renormalization schemes. Before going ahead we notice that 
$n_5$ and $\chi_5$ depend on the renormalization scheme used: this is not surprising because,
even if they correspond to physical observables, before renormalization they are divergent
and the three RSs, which formally correspond to different subtractions in the bare expressions
for $n_5$ and $\chi_5$,  lead to three different models
as we have summarized in Table~\ref{tab:1}.

\begin{table}[t!]
\begin{tabular}{||c|c|c|c||}
\hline
&RS1 & RS2 & RS3 \\
\hline
$\langle\sigma\rangle$ &$F_\pi$ & $F_\pi$ & 
$ F_\pi 
\left(
1 + a \mu_5^2/F_\pi^2
\right)$ \\
\hline
$m_\sigma^2$ &$M_\sigma^2$ & $M_\sigma^2 + b \mu_5^2$ &$M_\sigma^2 + c \mu_5^2$\\
\hline
\end{tabular}
\caption{\label{tab:1} $\langle\sigma\rangle$ and screening mass of the $\sigma$-meson 
at zero temperature within the three renormalization schemes discussed in the text.
The value of the positive constants $a$, $b$ and $c$ is not important for the purpose of the
Table and their exact values can be found in the main text. $F_\pi$ and $M_\sigma$
denote the pion decay constant and the $\sigma-$meson screening mass in the vacuum.}
\end{table}

\section{The Critical Temperature}
We now focus on the main goal of the present study, namely
to understand the effect of a renormalized vacuum term on the critical temperature
for chiral symmetry restoration at finite $\mu_5$.
Thanks to the analytical results discussed in the previous Section, and to the
expansion of $\Omega_T$ in Eq.~\eqref{eq:OmT} for large temperature,
we can observe in a transparent way the effect we look for.

The effect of the renormalization scheme on the critical temperature can be understood
by the computation of the second order Ginzburg-Landau coefficient.
For temperatures close to the critical temperature we write $\Omega = \beta_2\sigma^2$;
the coefficient $\beta_2$ is negative in the chirally broken phase and vanishes at the critical
temperature. Extracting the $O(\sigma^2)$ term from the thermodynamic potential we find
\begin{eqnarray}
\beta_2 &=& \beta_{2,0} + \beta_2(T), 
\label{eq:BETAtt}
\end{eqnarray}
where
\begin{eqnarray}
\beta_{2,0} &=& -\frac{\lambda F_\pi^2}{2} -\frac{g^4 N_c N_f}{8\pi^2}F_\pi^2
\nonumber\\
&&+g^2\mu_5^2\frac{N_c N_f}{2\pi^2}
\log\frac{\sigma}{F_\pi} \nonumber 
- \xi g^2 \mu_5^2\frac{N_c N_f}{2\pi^2}\nonumber\\
&&
\label{eq:b20}
\end{eqnarray}
and
\begin{eqnarray}
 \beta_2(T) &=& g^2 T^2\frac{N_c N_f}{12} \nonumber\\
 &&+g^2\mu_5^2\frac{N_c N_f}{2\pi^2}
 \left(
 \log\frac{\pi T}{\sigma} - \gamma_E - \frac{1}{2}
 \right),
 \label{eq:b2T}
\end{eqnarray}
where $\beta_{2,0}$ and $\beta_2(T)$ correspond to the vacuum and thermal fluctuations contributions
to $\beta_2$ respectively; for $\beta_2(T)$ we have used the high temperature expansion quoted in~\cite{Yu:2015hym}
paying attention to the different definition of the constituent quark mass; finally $\xi$ in Eq.~\eqref{eq:b20}
depends on the RS we use:
\begin{eqnarray}
\xi &=&0,~~~\text{RS1},\\
\xi &=&\frac{1}{2},~~~\text{RS2},\\
\xi &=&\frac{3}{2},~~~\text{RS3}.
\end{eqnarray}
 The log term in the above equations
exactly cancels out leaving a regular expression in the $\sigma\rightarrow 0$ limit.

Before commenting the effect of $\xi\neq0$ it is useful to remind that an inspection 
of Eqs.~\eqref{eq:b20} and~\eqref{eq:b2T} shows the competition of the vacuum and the
thermal fluctuations in order to determine the effect of $\mu_5$ on the critical temperature
at $\xi=0$.
As a matter of fact for $\sigma\rightarrow 0$ the vacuum term gives a negative contribution
to $\beta_2$ while the thermal part gives a positive contribution, hence favouring
restoration of chiral symmetry. When they sum up the thermal contribution wins the competition
and the net effect of $\beta_2$ is to lower the critical temperature.
For $\xi=0$ in RS1 the $O(\mu_5^2)$ the vacuum term does not add any further contribution 
to $\beta_2$ leading eventually to a decreasing critical temperature $\mu_5$
with increasing $\mu_5$. 

On the other hand when $\xi>0$ there is an additional
negative contribution    to $\beta_2$ due to the vacuum term, which in fact can change the overall sign
of the $O(\mu_5^2)$ leading to an increase of the critical temperature at finite $\mu_5$.
We can show this by an expansion of $\beta_2$ around 
the solution $T_c = T_c^0 + \delta T$ in powers of $\delta T$,
with $T_c^0$ being the critical temperature at $\mu_5=0$, and retaining only the lowest order in 
$\delta T$ and $\mu_5^2$. Taking into account that $\beta_2(T_c^0)=0$ at $\mu_5=0$ we find
\begin{equation}
\beta_2\approx\delta T\frac{g^2 N_c N_f}{6}T_c^0 
+\mu_5^2\frac{g^2 N_c N_f}{2\pi^2}
\left(
\log\frac{\pi T_c^0}{F_\pi} -\tilde\xi
\right)
\label{eq:betdeltaT}
\end{equation}
with $\tilde\xi = \xi + \gamma_E + 1/2$. The above equation shows that
the thermal part entering through the log-term, and the vacuum part
can compete at $\xi\neq0$; the overall sign of the $O(\mu_5^2)$ correction 
to $\beta_2$ depends on the numerical values of $T_c^0$ and $\xi$,
and for $\xi$ large enough the correction to $\beta_2$ is negative, hence $\mu_5$
leads to an increase of the critical temperature. 
By solving for $\delta T$ the equation $\beta_2=0$ we find
the critical temperature as a function of $\mu_5$, namely
\begin{equation}
T_c(\mu_5) = T_c^0 + \mu_5^2\frac{3}{\pi^2 T_c^0}
\left(
\xi + \gamma_E + \frac{1}{2} - \log\frac{\pi T_c^0}{F_\pi} 
\right),
\label{eq:betdeltaT2}
\end{equation}
which together with Eq.~\eqref{eq:betdeltaT} represents the main result of the present article.
In the above equation $T_c^0$ is the only free parameter once the renormalization scheme has been specified,
therefore we can predict the behaviour of $T_c(\mu_5)$ by varying $T_c^0$ within a reasonable range.

\begin{figure}[t!]
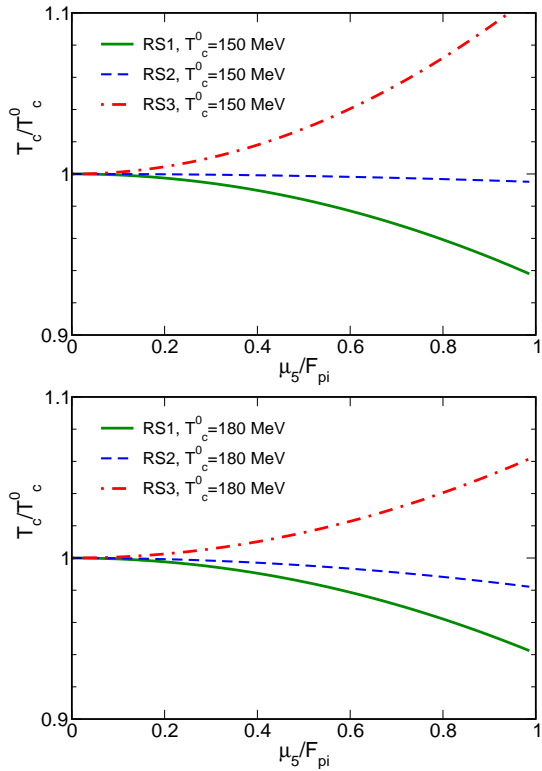

\begin{center}
\includegraphics[width=7cm]{figures/Tc.eps}\\
\includegraphics[width=7cm]{figures/Tc2.eps}
\caption{\label{Fig:1}Critical temperature as a function of $\mu_5$ for the three renormalization
schemes discussed in this article: upper panel corresponds to $T_c^0 = 150$ MeV and lower panel
to $T_c^0=180$ MeV. }
\end{center}
\end{figure}
In Fig.~\ref{Fig:1} we plot the critical temperature as a function of $\mu_5$ for the three renormalization
schemes discussed in this article: upper panel corresponds to $T_c^0 = 150$ MeV and lower panel
to $T_c^0=180$ MeV. The chiral chemical
potential leads to a lowering of $T_c$ within RS1 and RS2, in which the condensate at $T=0$
has no dependence on $\mu_5$. On the other hand within RS3 the critical temperature 
increases with $\mu_5$ for both values of $T_c^0$.
The results obtained within RS1 and RS2 are in agreement with previous model 
calculations~\cite{Gatto:2011wc,Fukushima:2010fe,Yu:2015hym,Chernodub:2011fr}
but in disagreement with recent lattice results~\cite{Braguta:2015owi,Braguta:2015zta}. 
On the other hand the scenario obtained within RS3 is in qualitative agreement with 
the latter data.

\section{Conclusions}
In this Article we have studied the catalysis of chiral symmetry breaking
due to a chiral chemical potential, $\mu_5$, within a renormalized quark-meson (QM) model.
The vacuum term at $\mu_5\neq0$ needs to be treated with care
because $\mu_5$ adds a log-type divergence to the vacuum energy which is transmitted
to physical quantities. In order to deal with ultraviolet divergences
we have introduced three different renormalization schemes (RSs) at $\mu_5\neq0$, whose
physical content is summarized in Table~\ref{tab:1} . 

We have then computed the critical temperature, $T_c$, for chiral symmetry restoration as a function $\mu_5$.
We have found that within RS1 and RS2 the critical temperature decreases with 
increasing $\mu_5$, in agreement with previous model calculations. 
On the other hand within RS3 we have found that $T_c(\mu_5)$ increases with $\mu_5$,
in agreement with recent lattice estimates of the same quantity as well as with previous analysis
based on universality \cite{Hanada:2011jb} and solution of Schwinger-Dyson equation \cite{Xu:2015vna}.

The reason of the discrepancy among the RSs is easy to understand.
Within the model at hand, as well as other simple chiral models,
the behaviour of $T_c(\mu_5)$ is understood as a competition between the vacuum energy on the
one hand, and on the thermal  part of the thermodynamic potential on the other 
hand~\cite{Gatto:2011wc,Yu:2015hym}. Expanding the thermodynamic potential
in power series of $\sigma$ in the vicinity of the phase transition, 
namely $\Omega = \beta_2\sigma^2$ with $\beta_2$ given by Equations~\eqref{eq:BETAtt}, 
\eqref{eq:b20} and~\eqref{eq:b2T}, it is clear that 
the vacuum term at $\mu_5\neq 0$ near the phase transition tends to lower the value of $\beta_2$, hence pushing 
the critical temperature towards larger values; on the other hand the contribution of the thermal fluctuations
to $\beta_2$ is positive, hence these tend to lower the critical temperature. 
In previous model calculations it has been always found that
eventually the competition is won by the thermal fluctuations because the vacuum term 
was not strong enough. We have obtained the same scenario within the RS1 and RS2.
On the other hand within RS3 the vacuum energy gets an extra negative term from renormalization
which makes it strong enough to compete against thermal fluctuations and 
push $T_c(\mu_5)$ towards larger values.
This is clearly represented by Eq.~\eqref{eq:betdeltaT2} in which $\xi$ depends on the
RS used: $\xi=0$, $1/2$ and $3/2$ for RS1, RS2 and RS3 respectively.

We would like to remark that even if $T_c$ represents a physical quantity in principle measurable,
in our calculations it depends explicitly on the RS because the three RSs introduced 
here correspond eventually to three different models, in which the  ground state at $T=0$ 
reacts in a different way to $\mu_5$.

The qualitative agreement of RS3 with lattice data is encouraging so we are tempted to take RS3
as the best RS among the ones studied here; however we should wait for the results
of lattice simulations with masses close to the physical ones before making such a statement,
and at the moment the most fair attitude is to consider the results of the present study as an exploration of the possible
scenarios that the QM model can predict.

In conclusion, the lesson we have learned from this study is that 
the vacuum energy in chiral models has to be treated with care; a similar 
call for a proper treatment of the vacuum energy was given in~\cite{Ruggieri:2013cya}
in which it was shown that the order of the phase transition in a magnetic field
can change  depending on the way the vacuum energy is
renormalized. In the context of strong interactions at finite $\mu_5$
we have found that predictions about the behaviour of the critical line at finite $\mu_5$ depend on the
way the divergence of the vacuum energy is treated. Moreover we have shown that
a simple chiral model like the QM model
is capable to reproduce qualitatively the recent lattice data about
$T_c(\mu_5)$, within a mean field approximation, and  without the addition of any exotic ingredient,
once a renormalization scheme is chosen.
The results presented here are thus encouraging and suggest to further investigate
properties of strongly interacting matter with chiral chemical potential
by the QM model. 

Finally we would like to mention that a possible future direction for investigation is the study of
in-medium properties of the system with finite chirality, using for example the strategies of
\cite{Peng:2002ie,Peng:2003jh,Peng:2004ev,Tang:2011za}. 
It is also worth to mention that in chiral models it is often assumed a 
local interaction which leads to a momentum-indepenent fermion self-energy at tree level,
in turn implying the consideration of all momentum modes
in the chiral condensate at finite temperature and these might affect the role of the thermal
fluctuations on the critical temperature.  In our opinion it is urgent to understand  the effect of 
a momentum-dependent quark self-energy on the evolution of the chiral condensate
at finite $\mu_5$ and $T$.
Works along these directions are already in progress and we plan to report on these topics in the near future.

{\em Acknowledgments}. The authors would like to thank the 
CAS President's International Fellowship Initiative (Grant No. 2015PM008), 
and the NSFC projects (11135011 and 11575190). 
M.~R. also acknowledges discussions with M. Frasca and M. Huang.


\begin{thebibliography}{99}

\bibitem{Adler:1969gk} 
  S.~L.~Adler,
  Phys.\ Rev.\  {\bf 177}, 2426 (1969).
  
\bibitem{Bell:1969ts} 
  J.~S.~Bell and R.~Jackiw,
  Nuovo Cim.\ A {\bf 60}, 47 (1969).

\bibitem{Moore:2000ara} 
  G.~D.~Moore,
  hep-ph/0009161.
\bibitem{Moore:2010jd} 
  G.~D.~Moore and M.~Tassler,
  JHEP {\bf 1102}, 105 (2011).

  
\bibitem{Vilenkin:1980fu} 
  A.~Vilenkin,
  Phys.\ Rev.\ D {\bf 22}, 3080 (1980).

\bibitem{Vilenkin:1982pn} 
  A.~Vilenkin and D.~A.~Leahy,
  Astrophys.\ J.\  {\bf 254}, 77 (1982).
  
\bibitem{Kharzeev:2007jp} 
  D.~E.~Kharzeev, L.~D.~McLerran and H.~J.~Warringa,
  Nucl.\ Phys.\ A {\bf 803}, 227 (2008).
  
\bibitem{Fukushima:2008xe} 
  K.~Fukushima, D.~E.~Kharzeev and H.~J.~Warringa,
  Phys.\ Rev.\ D {\bf 78}, 074033 (2008).
 
\bibitem{Kharzeev:2013ffa} 
  D.~E.~Kharzeev,
  Prog.\ Part.\ Nucl.\ Phys.\  {\bf 75}, 133 (2014).
\bibitem{Kharzeev:2015znc} 
  D.~E.~Kharzeev, J.~Liao, S.~A.~Voloshin and G.~Wang,
  arXiv:1511.04050 [hep-ph].
  
   
\bibitem{Son:2009tf} 
  D.~T.~Son and P.~Surowka,
  Phys.\ Rev.\ Lett.\  {\bf 103}, 191601 (2009).

\bibitem{Banerjee:2008th} 
  N.~Banerjee, J.~Bhattacharya, S.~Bhattacharyya, S.~Dutta, R.~Loganayagam and P.~Surowka,
  JHEP {\bf 1101}, 094 (2011).
  
\bibitem{Landsteiner:2011cp} 
  K.~Landsteiner, E.~Megias and F.~Pena-Benitez,
  Phys.\ Rev.\ Lett.\  {\bf 107}, 021601 (2011).
  
\bibitem{Son:2004tq} 
  D.~T.~Son and A.~R.~Zhitnitsky,
  Phys.\ Rev.\ D {\bf 70}, 074018 (2004).
  
\bibitem{Metlitski:2005pr} 
  M.~A.~Metlitski and A.~R.~Zhitnitsky,
  Phys.\ Rev.\ D {\bf 72}, 045011 (2005).
  
\bibitem{Kharzeev:2010gd} 
  D.~E.~Kharzeev and H.~U.~Yee,
  Phys.\ Rev.\ D {\bf 83}, 085007 (2011).
  
\bibitem{Chernodub:2015gxa} 
  M.~N.~Chernodub,
  JHEP {\bf 1601}, 100 (2016).
  
\bibitem{Chernodub:2015wxa}
  M.~N.~Chernodub and M.~Zubkov,
  arXiv:1508.03114 [cond-mat.mes-hall].
  
\bibitem{Chernodub:2013kya} 
  M.~N.~Chernodub, A.~Cortijo, A.~G.~Grushin, K.~Landsteiner and M.~A.~H.~Vozmediano,
  Phys.\ Rev.\ B {\bf 89}, no. 8, 081407 (2014).
  
\bibitem{Braguta:2013loa} 
  V.~Braguta, M.~N.~Chernodub, K.~Landsteiner, M.~I.~Polikarpov and M.~V.~Ulybyshev,
  Phys.\ Rev.\ D {\bf 88}, 071501 (2013).
  
\bibitem{Li:2014bha} 
  Q.~Li {\it et al.},
  arXiv:1412.6543 [cond-mat.str-el].
    


\bibitem{Gatto:2011wc} 
  R.~Gatto and M.~Ruggieri,
  Phys.\ Rev.\ D {\bf 85}, 054013 (2012).
  
\bibitem{Fukushima:2010fe} 
  K.~Fukushima, M.~Ruggieri and R.~Gatto,
  Phys.\ Rev.\ D {\bf 81}, 114031 (2010).
  
\bibitem{Chernodub:2011fr} 
  M.~N.~Chernodub and A.~S.~Nedelin,
  Phys.\ Rev.\ D {\bf 83}, 105008 (2011).
  
\bibitem{Ruggieri:2011xc} 
  M.~Ruggieri,
  Phys.\ Rev.\ D {\bf 84}, 014011 (2011).
 
\bibitem{Yu:2015hym} 
  L.~Yu, H.~Liu and M.~Huang,
  arXiv:1511.03073 [hep-ph].
  
\bibitem{Yu:2014xoa} 
  L.~Yu, J.~Van Doorsselaere and M.~Huang,
  Phys.\ Rev.\ D {\bf 91}, no. 7, 074011 (2015).
 
\bibitem{Braguta:2015owi} 
  V.~V.~Braguta, E.-M.~Ilgenfritz, A.~Y.~Kotov, B.~Petersson and S.~A.~Skinderev,
  arXiv:1512.05873 [hep-lat].
  
\bibitem{Braguta:2015zta} 
  V.~V.~Braguta, V.~A.~Goy, E.-M.~Ilgenfritz, A.~Y.~Kotov, A.~V.~Molochkov, M.~Muller-Preussker and B.~Petersson,
  JHEP {\bf 1506}, 094 (2015).
  
\bibitem{Braguta:2016aov} 
  V.~V.~Braguta and A.~Y.~Kotov,
  arXiv:1601.04957 [hep-th].
  
\bibitem{Hanada:2011jb} 
  M.~Hanada and N.~Yamamoto,
  PoS LATTICE {\bf 2011}, 221 (2011)
  [arXiv:1111.3391 [hep-lat]].
  
\bibitem{Xu:2015vna} 
  S.~S.~Xu, Z.~F.~Cui, B.~Wang, Y.~M.~Shi, Y.~C.~Yang and H.~S.~Zong,
  Phys.\ Rev.\ D {\bf 91}, no. 5, 056003 (2015).

\bibitem{Skokov:2010sf} 
  V.~Skokov, B.~Friman, E.~Nakano, K.~Redlich and B.-J.~Schaefer,
  Phys.\ Rev.\ D {\bf 82}, 034029 (2010).
  
\bibitem{Ruggieri:2013cya} 
  M.~Ruggieri, M.~Tachibana and V.~Greco,
  JHEP {\bf 1307}, 165 (2013).
  
  
\bibitem{Gervais:1969zz} 
  J.~L.~Gervais and B.~W.~Lee,
  Nucl.\ Phys.\ B {\bf 12}, 627 (1969).
\bibitem{Mota:1999hb} 
  A.~L.~Mota, M.~C.~Nemes, B.~Hiller and H.~Walliser,
  Nucl.\ Phys.\ A {\bf 652}, 73 (1999).
\bibitem{Jungnickel:1995fp} 
  D.~U.~Jungnickel and C.~Wetterich,
  Phys.\ Rev.\ D {\bf 53}, 5142 (1996).
\bibitem{Herbst:2010rf} 
  T.~K.~Herbst, J.~M.~Pawlowski and B.~J.~Schaefer,
  Phys.\ Lett.\ B {\bf 696}, 58 (2011).
\bibitem{Schaefer:2004en} 
  B.~J.~Schaefer and J.~Wambach,
  Nucl.\ Phys.\ A {\bf 757}, 479 (2005).
  
\bibitem{Suganuma:1990nn} 
  H.~Suganuma and T.~Tatsumi,
  Annals Phys.\  {\bf 208}, 470 (1991).
  
\bibitem{Frasca:2011zn} 
  M.~Frasca and M.~Ruggieri,
  Phys.\ Rev.\ D {\bf 83}, 094024 (2011).
  
\bibitem{Fukushima:2003fw} 
  K.~Fukushima,
  Phys.\ Lett.\ B {\bf 591}, 277 (2004)
  




\bibitem{Peng:2002ie} 
  G.~X.~Peng, H.~C.~Chiang, U.~Lombardo and M.~Loewe,
  Phys.\ Lett.\ B {\bf 548}, 189 (2002).
  
\bibitem{Peng:2003jh} 
  G.~X.~Peng, U.~Lombardo, M.~Loewe, H.~C.~Chiang and P.~Z.~Ning,
  Int.\ J.\ Mod.\ Phys.\ A {\bf 18}, 3151 (2003).
  
\bibitem{Peng:2004ev} 
  G.~X.~Peng,
  Nucl.\ Phys.\ A {\bf 747}, 75 (2005).
  
\bibitem{Tang:2011za} 
  H.~H.~Tang and G.~X.~Peng,
  Commun.\ Theor.\ Phys.\  {\bf 56}, 1071 (2011).
  
  \end{thebibliography}
\end{document}